# Inflation, unemployment, and labor force. Phillips curves and long-term projections for Japan

Oleg I. Kitov

University of Oxford

Ivan O. Kitov

Institute for the Geospheres' Dynamics, Russian Academy of Sciences

**Abstract**

The evolution of the rate of price inflation, $\pi(t)$, and unemployment, $u(t)$, in Japan has been modeled within the Phillips curve framework. As an extension to the Phillips curve, we represent both variables as linear functions of the change rate of labor force. All models were first estimated in 2005 for the period between 1980 and 2003. Here we update these original models with data through 2012. The revisited models accurately describe disinflation during the 1980s and 1990s as well as the whole deflationary period started in the late 1990s. The Phillips curve for Japan confirms the original concept that growing unemployment results in decreasing inflation. A linear and lagged generalized Phillips curve expressed as a link between inflation, unemployment, and labor force has been also re-estimated and validated by new data. Labor force projections allow a long-term inflation and unemployment forecast: the GDP deflator will be negative (between -0.5% and -2% per year) during the next 40 years. The rate fo unemployment will increase from ~4.3% in 2012 to ~5.5% in 2050.

**Introduction**

This paper revisits several empirical relationships between inflation, $\pi(t)$, unemployment, $u(t)$, the change rate of labor force level, $d\ln LF(t)/dt$, previously estimated for Japan within the Phillips curve framework. The original Phillips curve (PC) formulation (1958) as well as the reverse direction of causation introduced by Fisher (1926) considers the link between inflation and unemployment as a causal one – there were no autoregressive terms. Following these original assumptions behind the link between inflation and unemployment, we first estimated the causal relationship similar to the Phillips curve nine years ago (Kitov, 2006). Here its performance is tested using new data obtained since 2003 in order to validate our original findings. In addition to the standard PC and following the approach developed by Stock and Watson (1999), we took the advantage of pure econometric techniques and extended the set of economic activity by workforce. The rate of labor force change has never been tested in any econometric setting as a predictor of inflation and/or unemployment except those developed by the authors (*e.g.*, Kitov and Kitov, 2010). We have estimated two individual relationships between the change rate of *LF* and inflation, and the change rate of *LF* and unemployment as well as a generalized relationship between all three variables.

The re-estimated relationships confirm the whole set of original models. This allows answering important practical questions addressed in the studies of inflation and unemployment as related to the degree of control over these variables by monetary and fiscal authorities. The 2007 crisis and dramatic change in all parameters expressing economic activity (*e.g.*, the rate of unemployment in the US or real GDP per capital in Japan) without any significant change in the rate of price inflation put under doubt central bank's capability to affect inflation and unemployment when conducting monetary policy. Since the causal links validated in this paper do not change empirical coefficients and thus contain no structural breaks it is unlikely that the Bank of Japan has been controlling inflation and unemployment. Numerous labor force projections for Japan include a rapidly falling working age population (*e.g*., Shirakawa, 2012; Bullard *et al*. 2012) and decreasing rate of participation in labor force (*e.g.,* Kawata and Naganuma, 2010). Our model predicts that the shrinking labor force will induce a long-term deflation period stretching into the second half of the 21$^{st}$ century which will be accompanied by an elevated unemployment rate.

There are studies telling a slightly different story. Leigh (2004) examined the influence of monetary policy on the liquidity trap in Japan. Was there some monetary policy which the Bank of Japan could conduct to avoiding deflationary slump? He found that the trap arose not because of monetary policy mistakes and that "a policy of responding more aggressively to the inflation gap while keeping the low inflation target would have provided little improvement in economic performance". This conclusion is partly in line with our findings, which deny any possibility of the influence on inflation except that transmitted through inflation dependence on labor force and unemployment.

There exists a common opinion in the economists and central bankers community that inflation is a monetary phenomenon. Nelson (2006) investigated this assumption as applied to Germany and Japan and argued that the experiences of these countries in the 1970s indicate that once inflation is accepted by policymakers as a monetary phenomenon, the main obstacle to price stability has been overcome. Hence, central banks are able to control inflation through monetary policy.

De Veirman (2007) studied, within the New Keynesian Phillips Curve (NKPC) framework, the output-inflation trade-off in Japan as a linear relationship with a time-invariant slope during the period between 1998 and 2002. He found that large negative output gap did not cause accelerating deflation, which would be expected according to the NKPC. Kamada (2004) investigated the importance of various real-time measures of output gap for inflation prediction and development of monetary policy by the Bank of Japan and reported that some measures of output gap to be marginally useful for the inflation prediction despite problems with high uncertainty in real-time estimates. The Taylor rule needed more ingredients "for preventing the asset bubble". These findings do not contradict our model since deflation is a natural result of decreasing labor force level in Japan, not output gap.

Sekine (2001) studied inflation function and forecasts at a one-year horizon for Japan using equilibrium correction model. He demonstrated only marginal forecast improvement, relative to the simplest autoregressive (AR) model, even when such variables as markup relationships, excess money, and output gap are included. Feyzioğlu and Willard (2006) found that foreign countries, specifically China, have no influence on prices in Japan. These results also do not contradict the dependence of inflation on labor force.

The evolution of unemployment in Japan was also studied out of the PC framework. Caporale and Gil-Alana (2006) tested unemployment time series in Japan for structural breaks at unknown points. They showed that structuralist approach to unemployment works well in Japan and interpreted this observation using specific features of labor market. Only one structural break near 1993 was identified in the Japanese unemployment time series. Pascalau (2007) found a long-run equilibrium relation between unemployment rate, productivity, and real wages in Japan. All the involved variables had a unit root and, thus, cointegration tests with non-linear error-correction mechanisms were applied. The author reported relatively long persistency of unemployment shocks.

Kitov (2006) estimated empirical coefficients for various representations of the Phillips curve in Japan as based on the link between inflation, unemployment, and labor force change. Instead of standard econometric methods implying stochastic trends in non-stationary time series, a simplified 1-D version of the boundary element method (Kitov and Kitov, 2010) with the LSQ technique was applied to cumulative curves. For the rate of CPI inflation (with imputed rent) and labor force, the model had a slope of 1.77 and constant term -0.0035. It was also found that the change in labor force occurred practically simultaneously with that in inflation.

The remainder of this paper consists of three Sections and Conclusion. Section 1 discusses the Phillips curve family. Section 2 presents a revised PC for Japan, and Section 3 reports some quantitative results for two individual and the generalized link between all three involved variables. Labor force projections are used to predict the evolution of inflation and unemployment between 2010 and 2050.

### 1. The Phillips curve family

Unwinding the history of inflation models to their common root Gordon (2011) traced them to the seminal Phillips' paper (1958). The original Phillips curve implied a causal and nonlinear link between the rate of change of the nominal wage rate, $w_t$, and the contemporary rate of unemployment:

$$w_t = -0.90 + 9.64 u_t^{-1.39} \qquad (1)$$

Relationship (1) suggests that wages are driven by the change in unemployment rate. The assumption of a causal link worked well for some periods in the UK. When applied to inflation and unemployment measurements in the U.S., the PC successfully explained the 1950s. The PC became an indispensible part of macroeconomics which has been extensively used by central banks ever since. The success of (1) did not last long, however, and new data measured in the late 1960s and early 1970 challenged the original PC version. We follow up the original assumption of a causal link between inflation and unemployment to construct an empirical Phillips curve for Japan.

In the 1970s, the PC concept in the mainstream theory divided into two larger branches, both included autoregressive properties of inflation and unemployment. The latter was also replaced by different parameters of economic activity. The underlying assumption of the causal link was abandoned and replaced by the hypothesis of "rational expectations" (Lucas, 1972, 1973) and later by the concept of "inflation expectations" (*e.g.*, Gali and Gertler, 1999). The former approach includes a varying number of past inflation values (i.e. autoregressive terms). It was designed to explain inflation persistency during the high-inflation period started in the early 1970s and ended in the mid 1980s.

The concept of inflation expectations surfaced in the late 1990s in order to explain the Great Moderation as controlled by monetary and fiscal authorities (Sims, 2007, 2008). Due to the conceptual legacy, the term New Keynesian Phillips Curve is often used for the inflation expectation type models. Despite highly elaborated mathematics representation and significant increase in the model dimensionality (*i.e.* in the number of defining parameters) relative to the parsimonious Phillips curve, both approaches have not been successful in quantitative explanation and prediction of inflation and/or unemployment (*e.g.*, Rudd and Whelan, 2005ab).

A pure econometric approach was introduced by Stock and Watson (1999), who tested a large number of Phillips-curve-based models for predictive power using various parameters of activity (both individually and in aggregated form) instead of and together with unemployment. This purely econometric approach did not include deep economic speculations and was aimed at finding technically appropriate predictors. The principal component analysis (*e.g*., Stock and Watson, 2002) was a natural extension to the multi-predictor models and practically ignored any theoretical background. Under the principal component approach, the driving forces of inflation are essentially hidden.

Three decades before Phillips, Irving Fisher (1926) introduced an opposite direction of causation (in this sense, it is an anti-Phillips curve) and described the mechanism of price inflation driving the rate of unemployment. Fisher analyzed monthly data for a short period between 1915 and 1925 using inflation lags up to five months. Both time series were too short and noisy for robust statistical estimates of coefficients and lags in the anti-Phillips causal relationship. For the U.S., Kitov (2009) estimated an anti-Phillips curve using observations between 1965 and 2008 and found that the change of unemployment lags behind the change in inflation by 30 months. The 43-year period provides an excellent resolution of regression coefficients and the lag. This anti-Phillips relationship was successfully tested for cointegration and Granger causality. In any case, the 30-month lag implies the direction of causation.

The original Phillips curve for the UK and the anti-Phillips curve introduced by Fisher both provide solid evidences for the existence of a causal link between inflation and unemployment. The conflict between the directions of causation can be resolved when both variables are driven by a third force with different lags. Depending on which lag is larger inflation may lag behind or lead unemployment. Co-movement found in Japan is just a degenerate case.

The framework of our study is similar to that introduced and then developed by Stock and Watson (2006, 2007, 2008) for many predictors. They assessed the performance of inflation forecasting in various specifications of the Phillips curve. Their study was partially initiated by the superior forecasting result of a univariate model (naïve prediction) demonstrated by Atkeson and Ohanian (2001). Stock and Watson convincingly demonstrated that neither before the 2007 crisis (2007) nor after the crisis (2010) can the Phillips curve specifications provide long term improvement on the naïve prediction at a one-year horizon.

Following Phillips and Fisher, we exclude autoregressive components from the Phillips curve and estimate two different specifications for inflation:

$$\pi(t) = \alpha + \beta u(t-t_0) + \varepsilon(t) \qquad (2)$$

$$\pi(t) = \alpha_1 + \beta_1 l(t-t_1) + \varepsilon_1(t) \qquad (3)$$

where $\pi(t)$ is the rate of price inflation at time $t$, $\alpha$ and $\beta$ are empirical coefficients of the Phillips curve with the time lag $t_0$, which can be positive or negative, $\varepsilon(t)$ is the error term, which we minimize by the LSQ for the cumulative curves, with the initial and final levels fixed to the observed ones; $l(t)=dlnLF(t)/dt$ is the rate of change in labor force, $\alpha_1$ and $\beta_1$ are empirical coefficients of the link between inflation and labor force, $t_1$ is the non-negative time lag of inflation, and $\varepsilon_1(t)$ is the model error.

Then, we represent unemployment as a linear and lagged function of the change rate in labor force:

$$u(t) = \alpha_2 + \beta_2 l(t-t_2) + \varepsilon_2(t) \qquad (4)$$

with the same meaning of the coefficients and the lag as in (3). We finalize the set of causal models with a generalized version:

$$\pi(t) = \alpha_3 + \beta_3 l(t-t_1) + \gamma_3 u(t-t_0) + \varepsilon_3(t) \qquad (5)$$

Relationships (2) through (5) have been carefully re-estimated with the data for the past nine years and compared to those obtained by Kitov (2006).

## 2. The Japan Phillips curve

We start with data evaluation for a conventional Phillips curve. Japan is a country with a modern statistical service. We borrowed all necessary time series of inflation and unemployment from the Statistics Bureau of the Ministry of Internal Affairs and Communications (SB, 2013), which also provides information on various economic and demographic variables. Alternatively, similar data sets are available from the U.S. Bureau of Labor Statistics (BLS, 2013) and from the Organization of Economic Cooperation and Development (OECD, 2013). Some of these sets are longer than those from the Japan SB.

For reliability of quantitative analysis, the most important issue is the quality of corresponding measurements. There are two main requirements to these data: they have to be as precise as possible in respect to any given definition, and the data must by comparable over time. The precision is related to methodology of measurements and implementation of corresponding procedures. The comparability is provided by the consistency of definitions and methodology. For example, the OECD (2005) provides the following information on the comparability of labor force and unemployment time series for Japan:

***Series breaks:** In 1967 the "household interview" method was replaced by the "filled-in-by-household" method and the survey questionnaire was revised accordingly.*

According to this statement one should not expect any breaks in (4) after 1967. Caporale and Gil-Alana (2006) found a clear structural break in 1993, but Hayashi (2005) did not report any break in the unemployment time series between 1960 and 1999. The CPI and DGDP time series in Japan have a break near 1973 (Hayashi, 2005; Kitov and Kitov, 2012). This study demonstrates that the Phillips curve for Japan has a break near 1982, which indicates the presence of some other problems in the general comparability of the measurements before and after 1982 or with structural breaks in regional CPI time series (Ikena, 2012).

There are several measures for price inflation. The most popular definitions for the overall price change are the GDP deflator, DGDP, and Consumer Price Index, CPI. Various inflation time series might be studied, but only two of them are used in this paper. Figure 1 shows two inflation estimates: the OECD GDP deflator and the CPI estimates provided by Japan Statistical Bureau (SB). Both series start in 1970 and thus provide the estimates of rate since 1971. (The OECD provides CPI estimates since 1960 but they are not corroborated by the SB.) The difference between the curves is illustrative. The GDP deflator curve is below that for the CPI inflation since 1995. One has to bear in mind that the latter variable is a larger part of the former one. For further quantitative analysis it is important to notice that the accuracy of annual estimates of CPI inflation in Japan is under doubt, as discussed by Shiratsuka (1999) and Ariga and Matsui (2002).

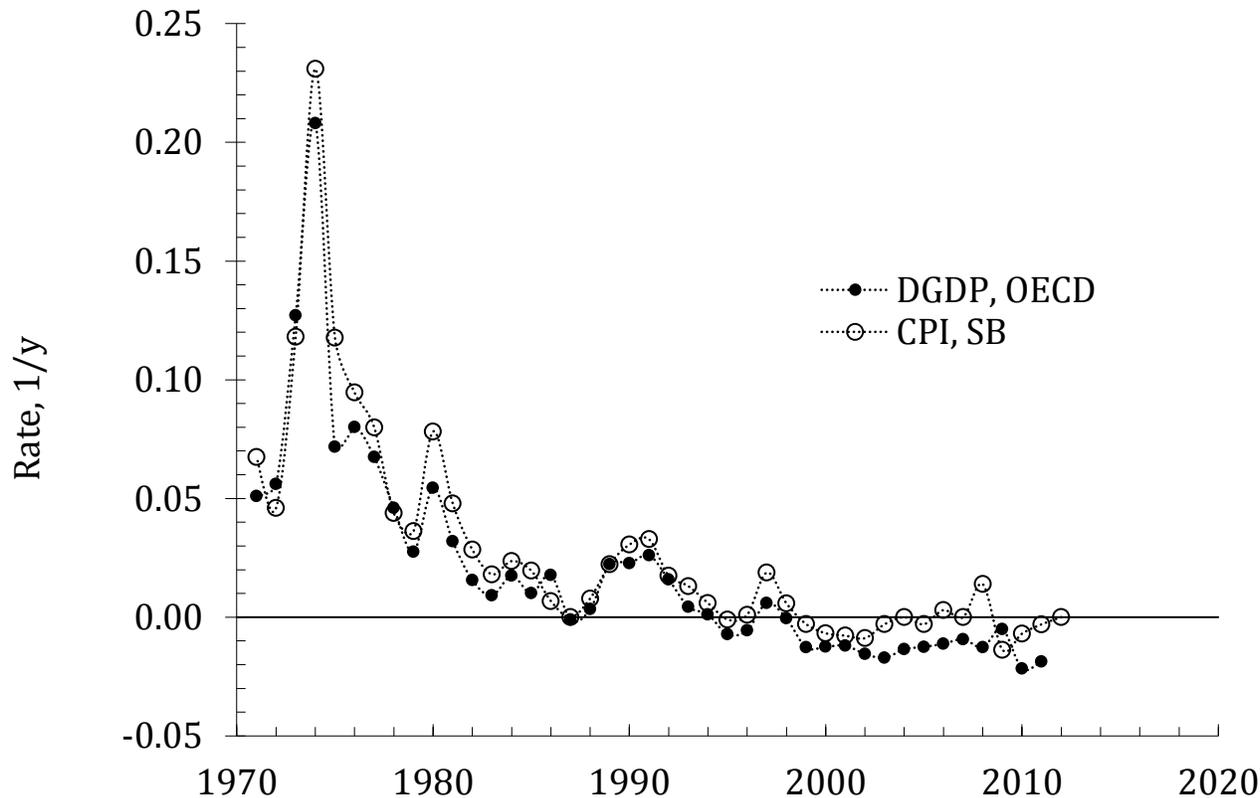

Figure 1. Comparison of two measures of inflation: CPI and GDP deflator. The curves are strikingly different during the whole period of measurements. The GDP deflator curve is below that for CPI after 1995 and before 1984. The gap between the curves has been growing with time.

We use two different estimates of unemployment in Japan provided by the Statistics Bureau and by the OECD. Figure 2 demonstrates that they are close and almost undistinguishable before 2005. True unemployment, as related to some perfect (but not currently available) definition of unemployment, might be between these curves and out of the curves as well. At the same time, both presented measures of unemployment are similar and it is likely that the true unemployment accurately repeats their shape. In this case, any of the measures can be used in quantitative modeling as representing the same portion of the true unemployment. Similar statement is valid for inflation measures. Apparently, actual problems are associated not with the difference between measured and true variables but with sudden jumps in the definitions of measured variables.

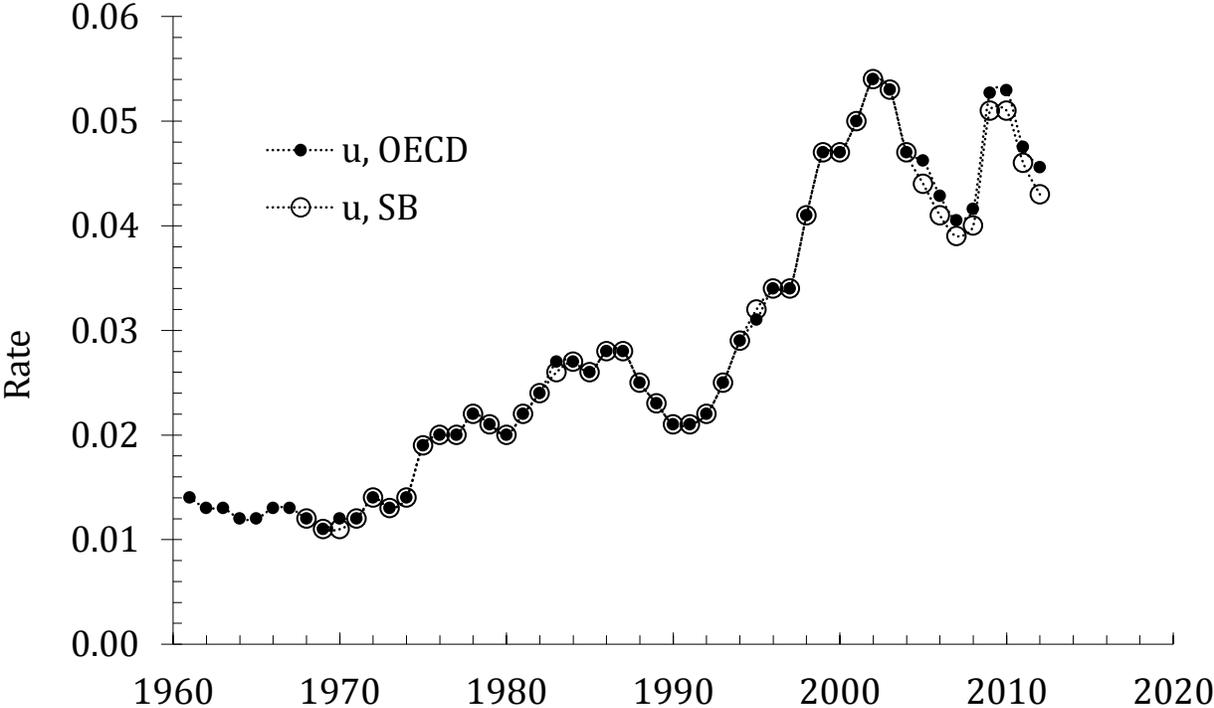

Figure 2. Comparison of two definitions of unemployment. The curves are slightly different since 2005.

Figure 3 presents a scatter plot for the rate of unemployment (SB) and CPI inflation in Japan since 1982. Linear regression gives reliable estimates of the slope of -0.93[0.11] and constant term 0.041[0.0043], with *p*-values $2.1 \cdot 10^{-10}$ and $4.3 \cdot 10^{-9}$, respectively. This regression has been calculated with various time shifts between the unemployment and inflation time series.

The best fit with $R^2$=0.70 was obtained when the unemployment curve and the inflation curve are not shifted. This is a standard Phillips curve, whatever the direction of causation is.

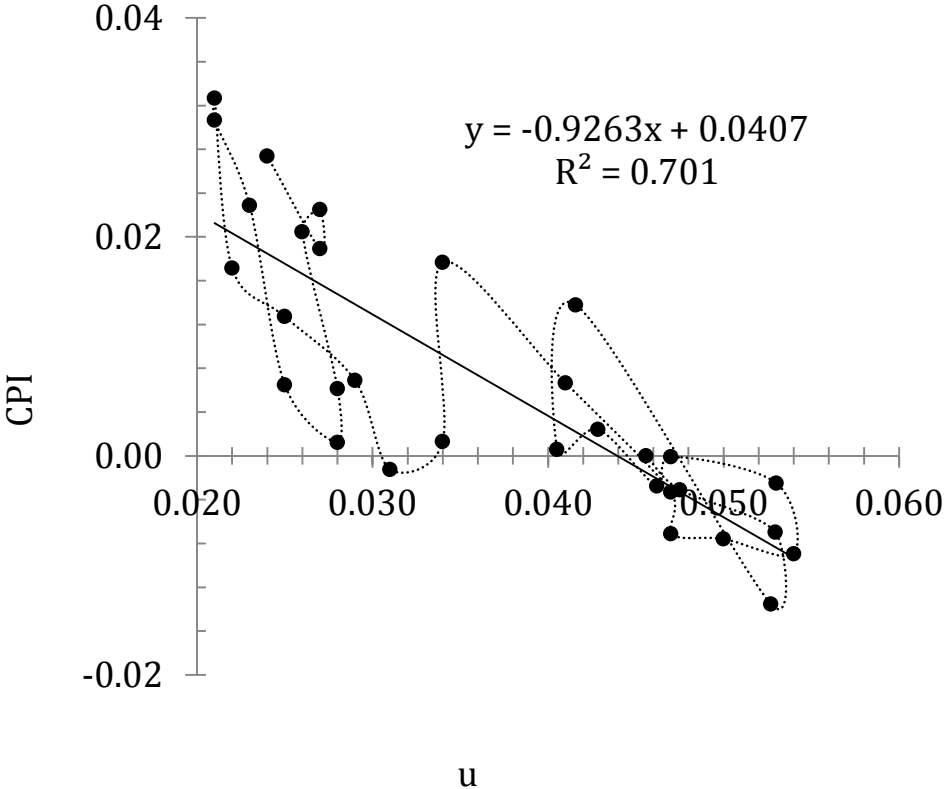

Figure 3. Inflation/unemployment scatter plot and linear regression for the period between 1982 and 2012. Neighboring years are connected by dotted curve. This is a standard representation of the Phillips curve for Japan. Regression coefficients are -0.93 and 0.041.

Fisher's representation of the (anti-)Phillips curve is depicted in Figure 4 together with the relevant error term. The predicted rate of unemployment demonstrates a relatively good agreement with the measured one since 1982 and the best-fit relationship is as follows:

$$u(t) = -1.10[0.14]\pi(t) + 0.044[0.005] ; \quad t > 1981 \tag{6}$$

The period after 2003 is highlighted by red triangles. The newly measured nine values from 2004 to 2012 are accurately predicted by the equation obtained for the previous period. This is an excellent validation of the original anti-Phillips curve for Japan. The error term looks random and does not vary in amplitude over time, but it is likely too short (31 readings) for a reliable unit root test. Standard deviation of the model error from 1982 to 2012 is ($\sigma=$) 0.007. Statistical estimates of the coefficients ($p$-value less than $10^{-9}$ for both coefficients) show high reliability of the anti-Phillips curve. Therefore, one can expect that decreasing inflation in the years to come will be accompanied by increasing unemployment.

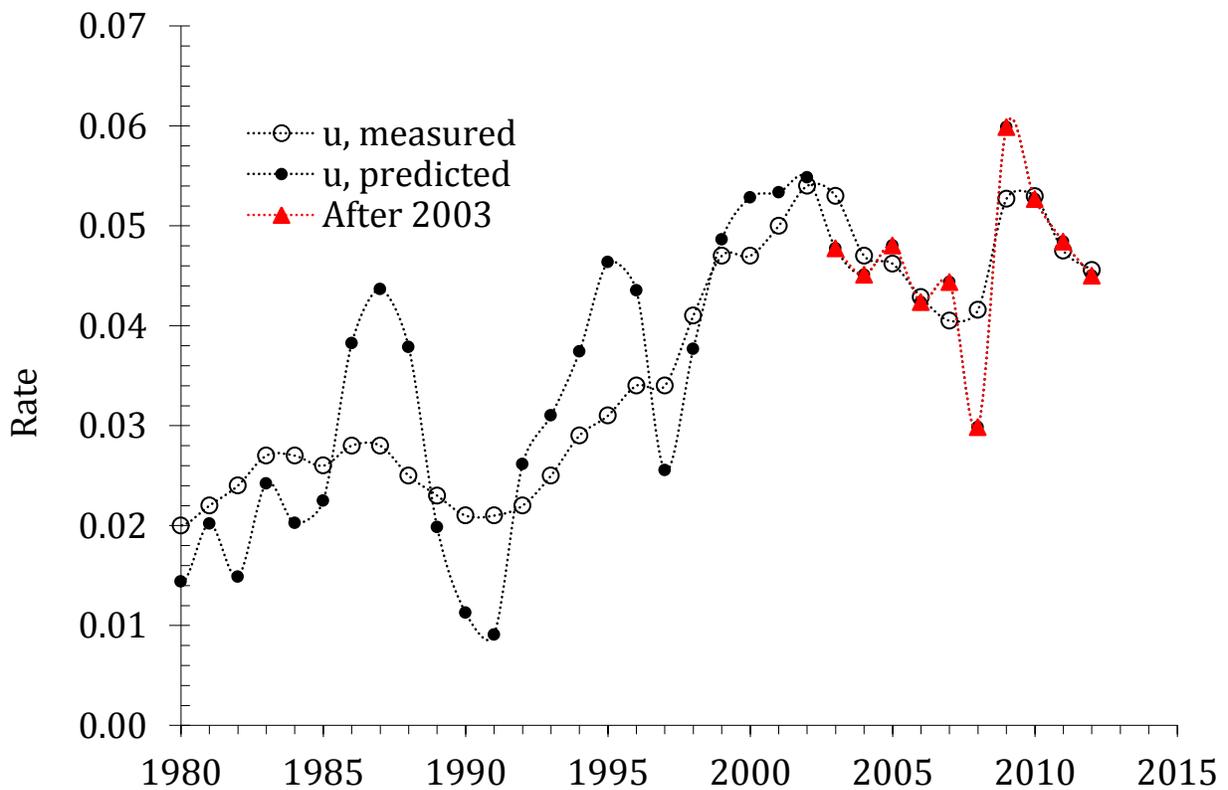

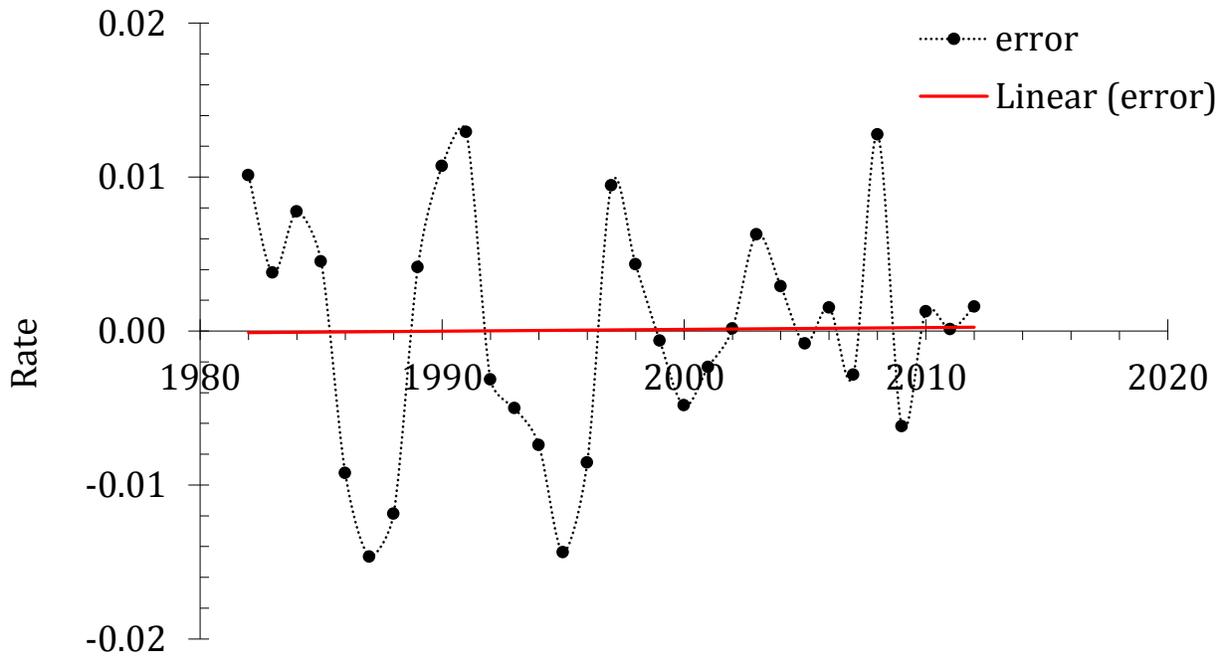

Figure 4. Upper panel: Measured unemployment and that predicted from CPI inflation for the period between 1982 and 2012. The coefficient of determination is the same as in Figure 3 and shows that 70% of variability in unemployment is explained by inflation. Lower panel: Error term with a regression line. Standard deviation is 0.007 for the period between 1982 and 2012.

The existence of a reliable Phillips curve in Japan raises a question about the consistency of monetary policy of the Bank of Japan. Does the bank conduct a monetary policy, which balances inflation and unemployment? Unlike Germany, where the Bundesbank has been showing during the last twenty five years the unwillingness to reduce unemployment in exchange for higher inflation, the BoJ was not able to decrease unemployment in order to get positive inflation figures. The next Section presents the driving force of both inflation and unemployment. Certainly, when controlled, the change in labor force might be the mechanism for monetary and fiscal authorities to fulfill their mandates.

## 3. Modeling inflation and unemployment in Japan

As many economic parameters, labor force estimates are also agency dependent due to various definitions and population adjustments. Figure 5 compares the change rate of labor force provided by the OECD (2013), Japan Statistics Bureau (SB, 2013), and BLS (2013). Despite strong similarity, some discrepancy reaching 0.01 (or 10% of the total labor force) is observed. Such a difference is an important indicator of the difficulties in labor force definition. Further investigations are necessary to elaborate a consistent understanding of the term "labor force". The model linking labor force change and inflation is likely a good candidate for quantitative consolidation of various definitions and approaches.

First, we test the existence of a link between inflation and labor force, which corresponds to a broader definition of Phillips curve. Because of the structural break in the 1980s, as described in the previous Section, we have chosen the period after 1982 for analysis. Varying time lag between labor force and inflation time series one can obtain the best-fit coefficients for the predicted CPI inflation, $\pi(t)$, according to the relationship:

$$\pi(t) = 0.0007[0.002] + 1.31[0.19]l(t-t_1) \tag{7}$$

where $t_1=0$ years. Kitov (2006) obtained slightly different coefficients for the period between 1981 and 2003, but these differences are negligible. Due to the shortness of the modeled period in the previous study, the estimate of coefficient $\beta_1$ was not very reliable. With nine new readings both coefficients are reliable. There is no time lag between inflation and unemployment in Japan. Coefficient $\alpha_1$, defining the level of inflation in the absence of labor force change, practically is undistinguishable from zero.

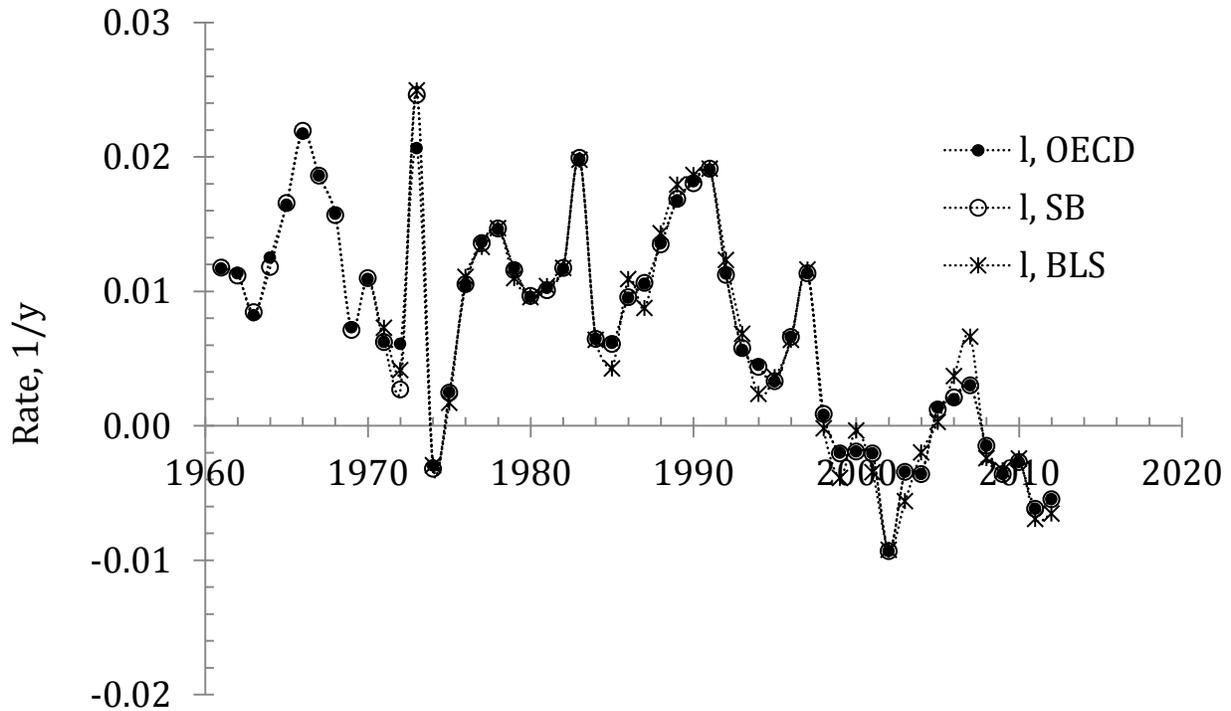

Figure 5. Comparison of three versions of the change rate of labor force level in Japan: OECD, Japan Statistics Bureau (SB), and the Bureau of Labor Statistics (BLS).

To estimate both coefficients in (7) we have applied a more precise and reliable technique based on the boundary element (or boundary integral) method (Kitov and Kitov, 2010), which is successfully used in physics and engineering. Suggesting a linear link between the rate of inflation and the change rate in labor force, both given as time derivatives, one can integrate linear differential equation (7). Having a Green's function for (7) and empirically estimated initial (boundary) conditions (zero and the cumulative inflation at time $t$) the BEM recommends applying the least squares fitting to estimate both coefficients. Because of the tight control on the integral model error, the BEM is superior to standard regression algorithms when the functional dependence in the Green's function is correct. The nine new readings should validate both the functional dependence and coefficients estimated for Japan in our previous study.

If the link between labor force and inflation is a causal one (similar to the Phillips curve or Fisher's representation), all short-term oscillations and uncorrelated noise in data should be related to inaccurate measurements. Any deviation induced by a force other than the labor force change (exogenous shock) will persist in the cumulative curves and destroy (7). (Similarly, the BEM is often based on conservation laws which do not allow any external forces to change invariant values carefully retained through the numerical solution.) The summation of the nonzero annual values of two RHS terms in (7) over 30 to 50 years make the predicted cumulative values very sensitive to both coefficients in (7), and thus, provide their extremely accurate estimates.

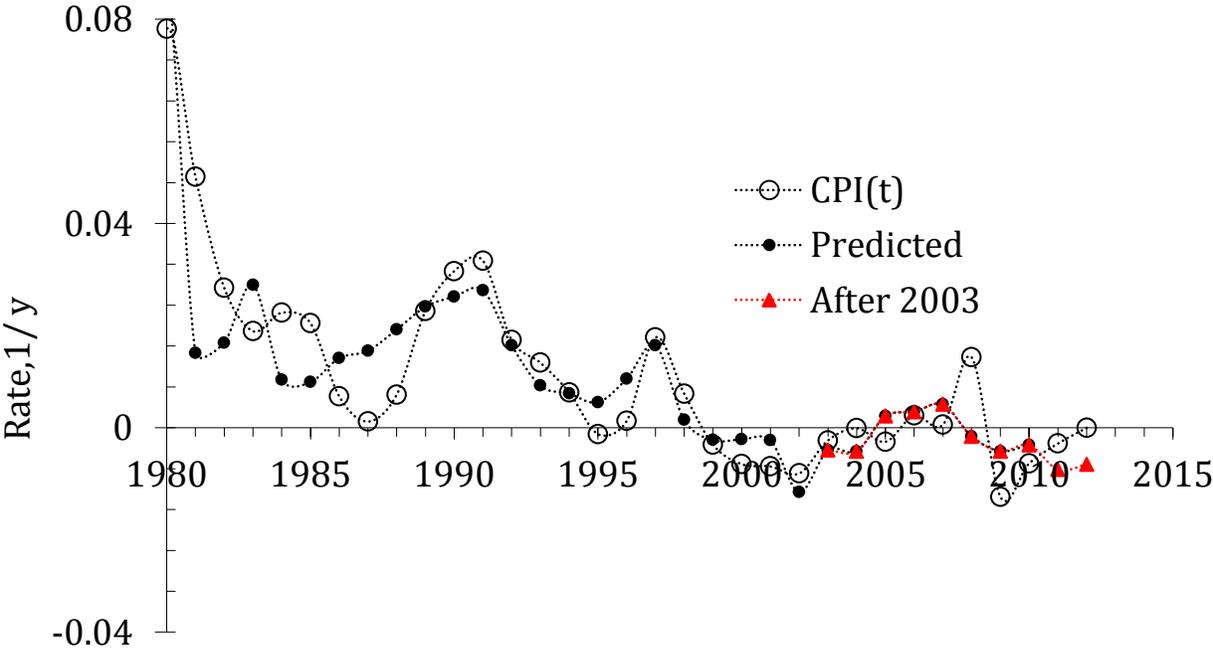

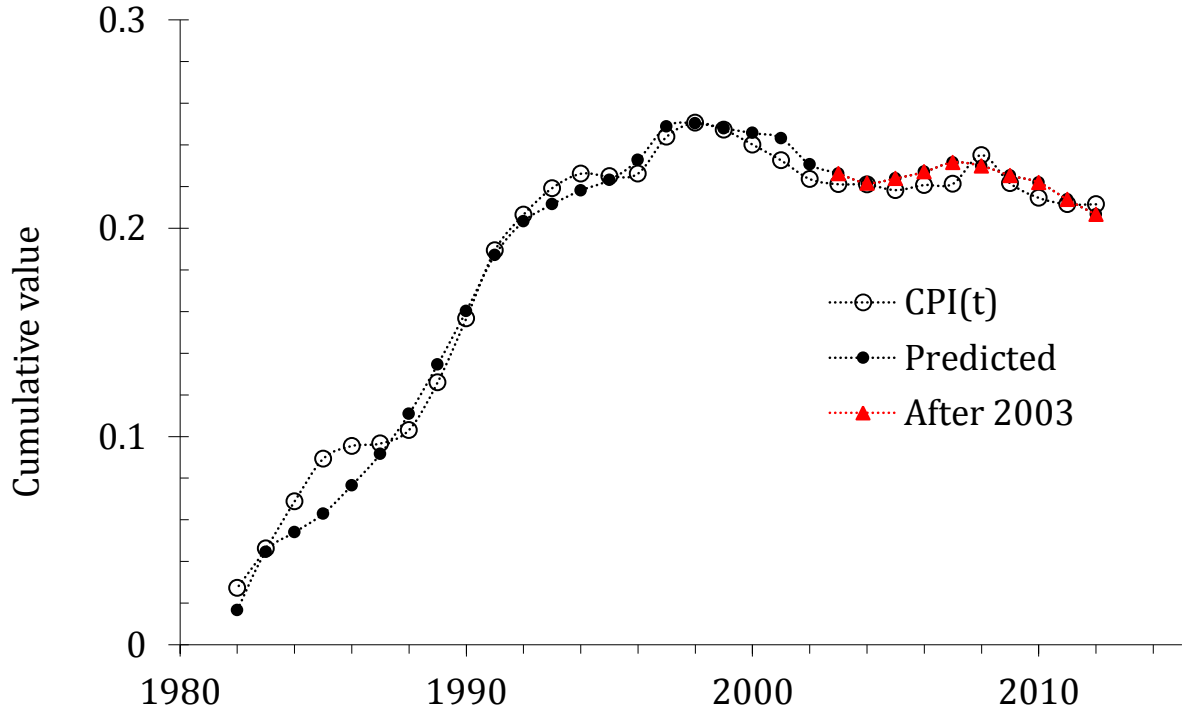

Figure 6. Upper panel: Measured inflation (CPI) and that predicted from the labor force change rate. Lower panel: Cumulative measured and predicted inflation curves used to estimate coefficients in (7).

The cumulative curves in Figure 6 are characterized by complex shapes. Before 1998, there was a period of intensive price growth ended with a deflationary period. The labor force change, defining the predicted inflation curve, follows all the turns in the measured cumulative inflation. One can conclude that relationship (7) is valid and the labor force change is the driving force of inflation.

Using the BEM, we have also estimated the best-fit coefficients for the predicted inflation expressed by the GDP deflator. The DGDP model differs from that for the CPI:

$$\pi(t) = -0.0084 + 1.90 l(t-t_1) \qquad (8)$$

where $t_1$=0 years. The estimated relationship has a different slope but a similar intercept term close to zero. This means that the watershed between inflation and deflation in Japan is a zero growth rate of labor force. Figure 7 displays the annual and cumulative curves both emphasizing the prediction for the period after 2003. The annual curves are smoothed with MA(3) in order to demonstrate that the model error is likely a realization of random noise – the predicted and observed curves practically coincide in amplitude and turning points. The predicted cumulative curve represents the total change in price inflation between 1981 (the initial value is 0) and 2012, but actually manifests the evolution of labor force experience by Japan during this period. The coefficient of determination for the cumulative curves is 0.99, with the model error (the difference between the cumulative curves) being an I(0) process. Since 1998, the absolute level of labor force has been falling and this is a long-term trend.

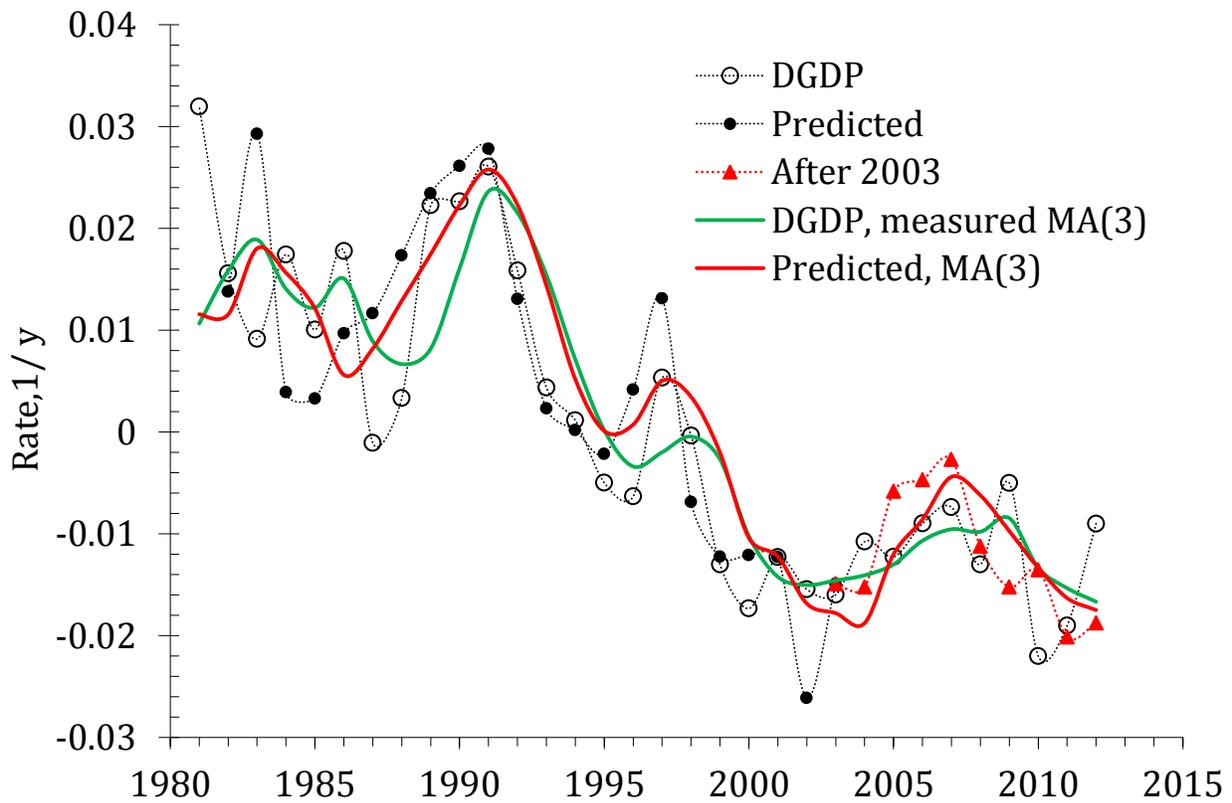

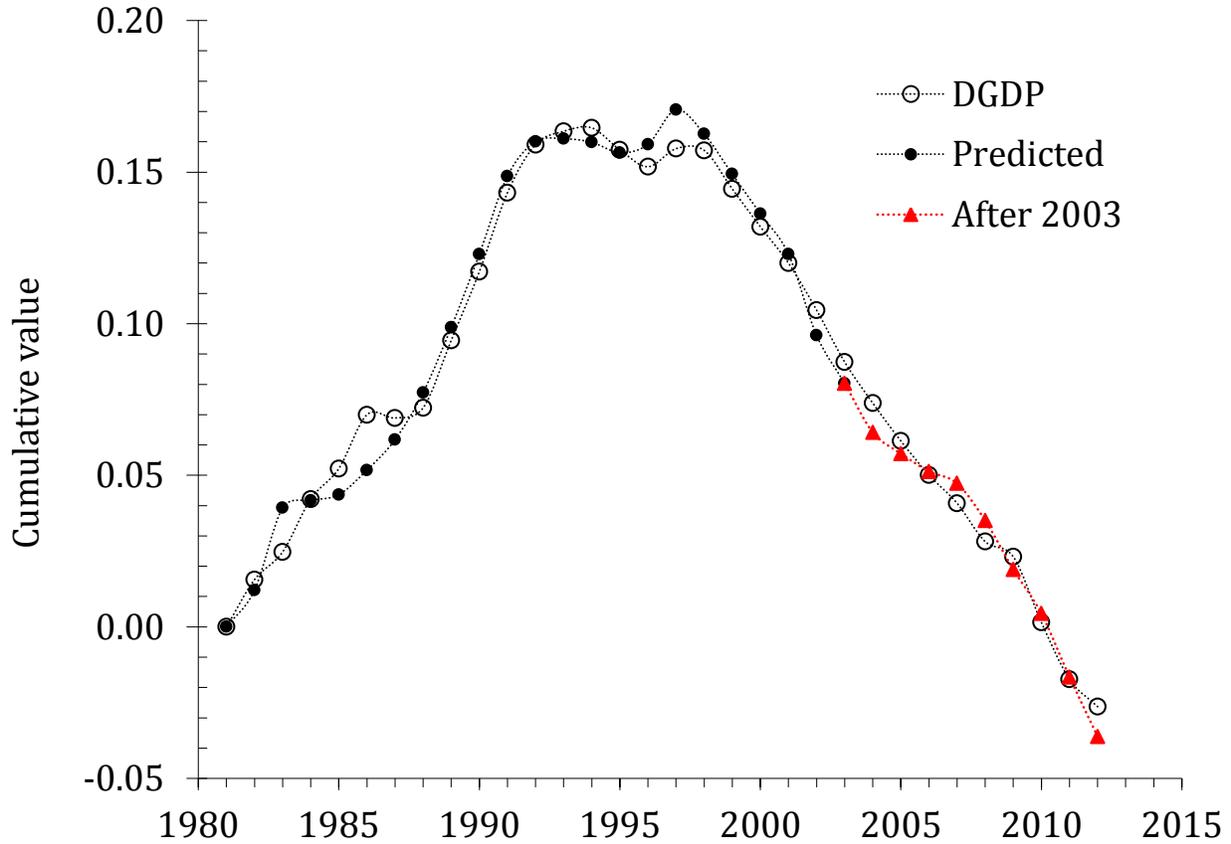

Figure 7. Upper panel: Measured inflation (DGDP) and that predicted from the labor force change rate. Both variables are smoothed with MA(3). Lower panel: Cumulative measured and predicted inflation curves used to estimate coefficients in (8).

It is difficult to precisely estimate the change in labor force level during one year. The coverage by labor force surveys is limited, definitions of labor force are not perfect, and population controls vary with time; especially, after censuses. As a remedy, special benchmarking procedures are implemented when all previous estimates are revised in order to match better measured level of labor force (*e.g.*, censuses). Therefore, the measuring precision of the change in labor force level should increase with time baseline. The net change during 30 years should be measured with lower relative uncertainty than during one year.

Now, we model the rate of unemployment as a function of labor force change. In addition to the use of BEM technique for coefficients in (4) we have introduced a structural break in 1976. The resulting relationship is as follows:

$$u(t) = -1.556 l(t) + 0.0432 \,;\quad t \geq 1977$$
$$u(t) = -0.179 l(t) + 0.0432 \,;\quad t < 1977 \tag{9}$$

The slope falls by an order of magnitude in 1977 with the intercept term not varying in time. Essentially, the period before 1977 is charaterized by a constant rate unepmloyment near 1%. Such a low and constant value may be related to specific definiton of unemployment.

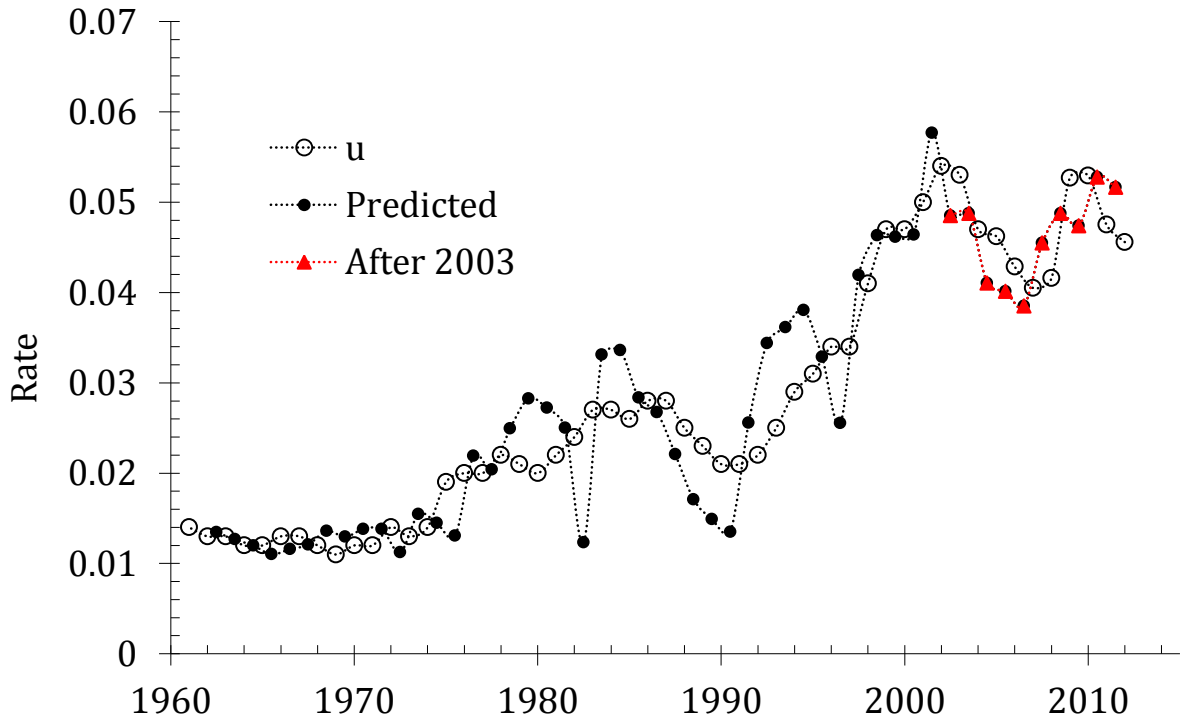

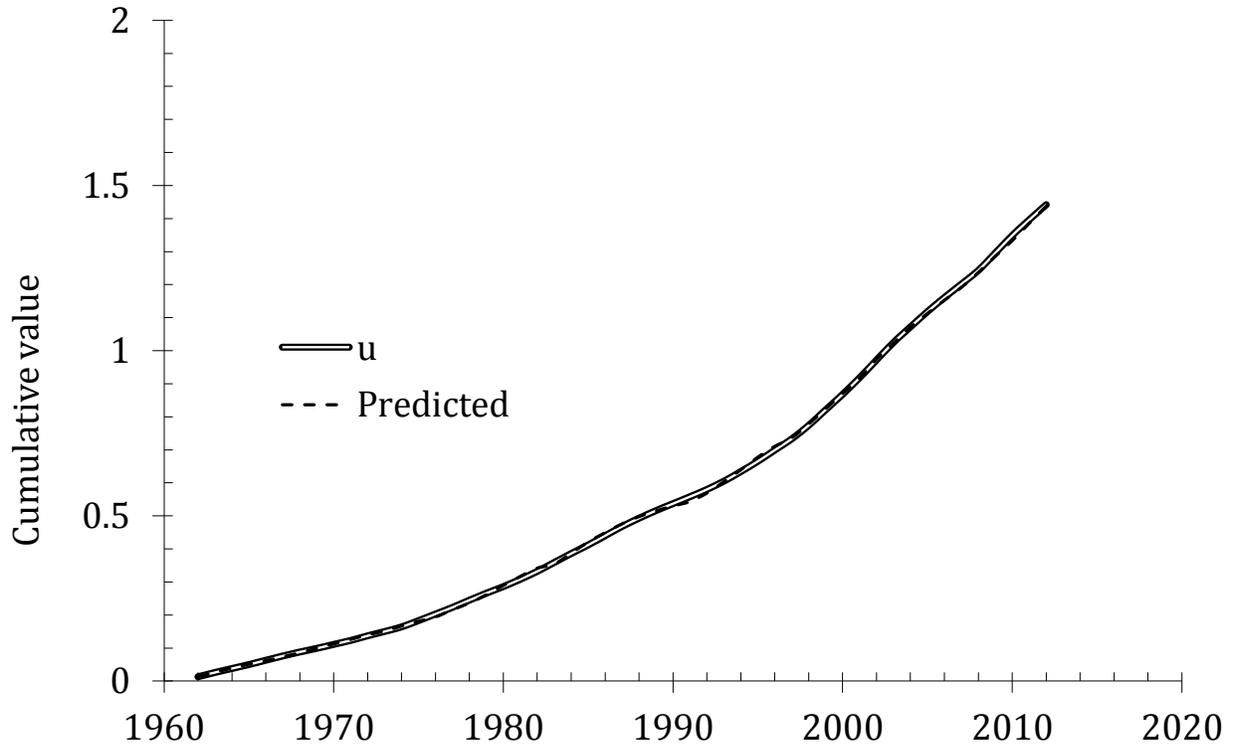

Figure 8. Comparison of measured unemployment and that predicted from the labor force change rate. Coefficients are obtained by the BEM. Upper panel: Annual readings. Lower panel: Cumulative curves.

Figure 8 displays annual and cumulative curves for the measured and predicted rate of unemployment between 1962 and 2012. The period between 2003 and 2012 is well described by the predicted time series as based on the coefficient estimated nine years ago. The cumulative curves are almost identical. An important feature of (9) is the negative relation between unemployment and labor force. Since 1962, increasing labor force has been associated with decreasing unemployment. Such a trade-off provides a useful tool to treat high unemployment – one needs to increase labor force. An again, the driving force of unemployment is the change in labor force, which is defined by the working age population and the rate of participation in labor force. With ageing population in Japan, both components will suffer a long term decline.

In the final part of our modeling, we gather three individual relationships in one generalized relation. Using the BEM and the assumption of a structural break, we have found the best-fit coefficients for the generalized equation in a Phillips curve representation, i.e. describing inflation as a function of two economic activity parameters:

$$\pi(t) = 2.80l(t) + 0.9u(t) - 0.0392; \quad t \geq 1982$$

$$\pi(t) = -10.0l(t) + 0.9u(t) + 0.161; \quad t < 1982 \tag{10}$$

Figure 9 compares the measured DGDP inflation reported by the OECD with the predicted from the OECD labor force and unemployment. The evolution of cumulative curves (lower panel) is very close. Therefore, the three involved variables are linked by a long term causal relation. Formally, one can carry out a cointegration test, which is applicable to stochastic time series, however. The Johansen test shows one cointegartion relation for these three variables. All unit root tests applied to the residual time series (model error) reject the null of a unit root. This also implies cointegration between the studied variables.

Having relationships (7) through (9), one can easily predict the evolution of inflation and unemployment between 2012 and 2050 using various labor force projections. The National Institute of Population and Social Security Research (http://www.ipss.go.jp) provides quantitative projections of total population, which can be used for labor force projection. We use the 2005 population projection and consider a constant rate of labor force participation fixed to 0.521, as measured in 2000. Figure 10 demonstrates that it was expected in 2005 that the level of labor force in Japan would decrease from 67,000,000 in 2010 to 57,000,000 in 2050. In reality, this old projection of labor force was too optimistic and the actual level is ~700,000 below the expected level. Nevertheless, we can use this projection as a conservative estimate of the future labor force in Japan.

Using the 2005 labor force projection we have estimated the future rate of inflation (DGDP) and unemployment. Figure 11 displays these predictions for the period through 2050. According to this labor force projection, 2007 was the last year of positive GDP deflator. Japan steps into a very long period of deflation. This economically negative process will be accompanied by increasing rate of unemployment. In the long run, unemployment will hover around 5.5% and the rate of deflation will reach 2.0% per year in 2050.

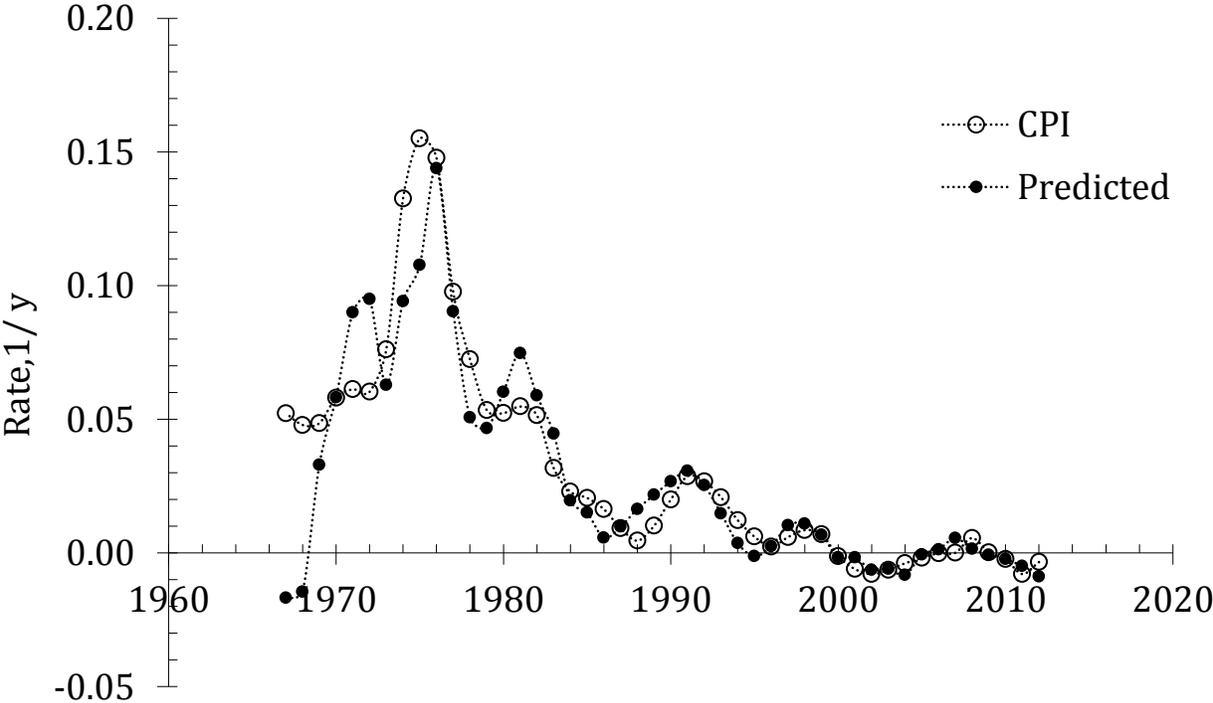

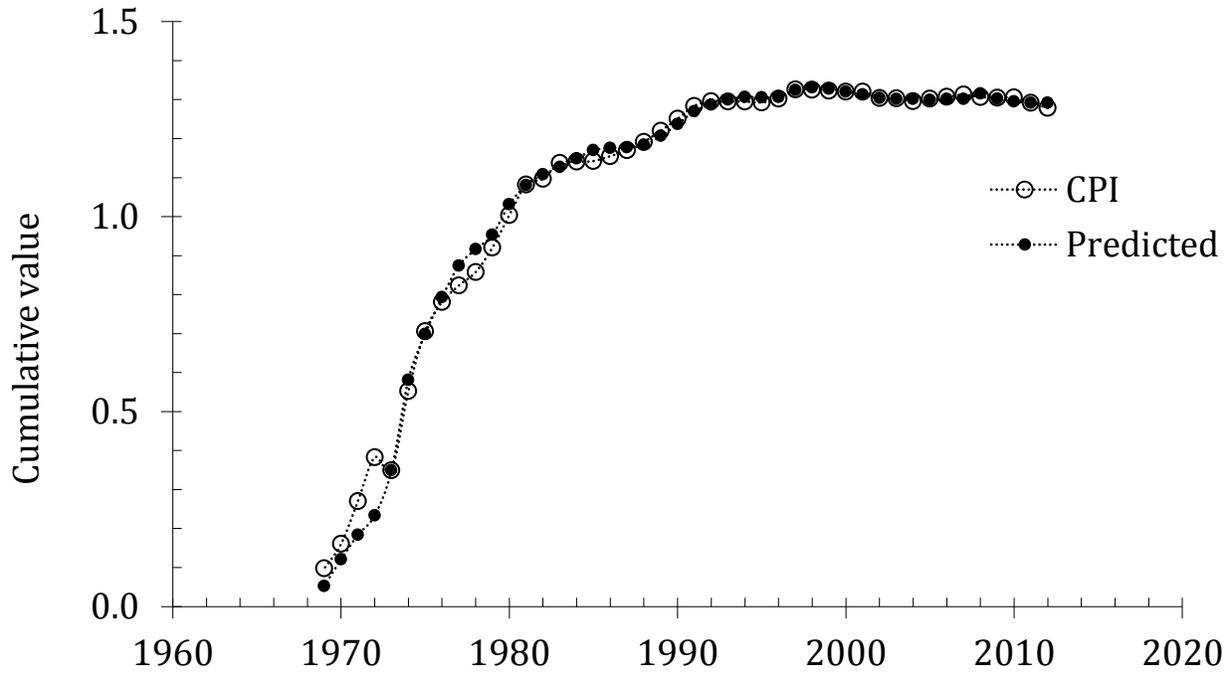

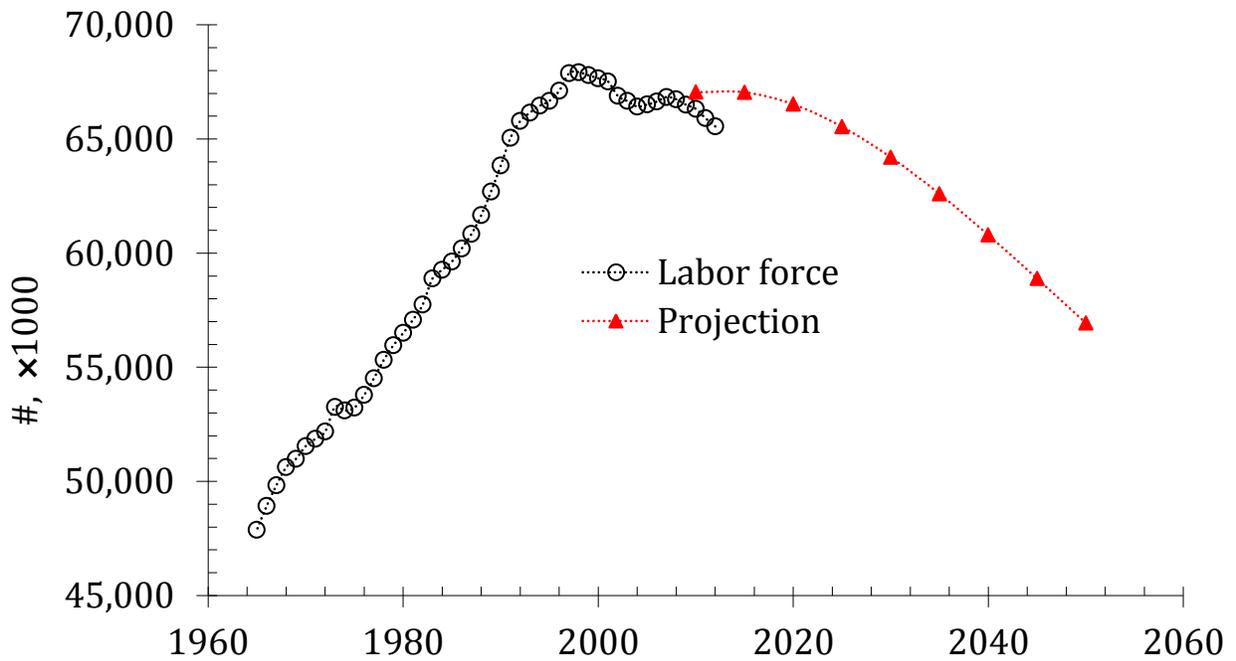

Figure 9. Upper panel: Measured inflation (CPI) and that predicted from the labor force change rate. Lower panel: Cumulative measured and predicted inflation curves.

Figure 10. Measured labor force between 1965 and 2012 and the 2005 projection of the labor force evolution through 2050.

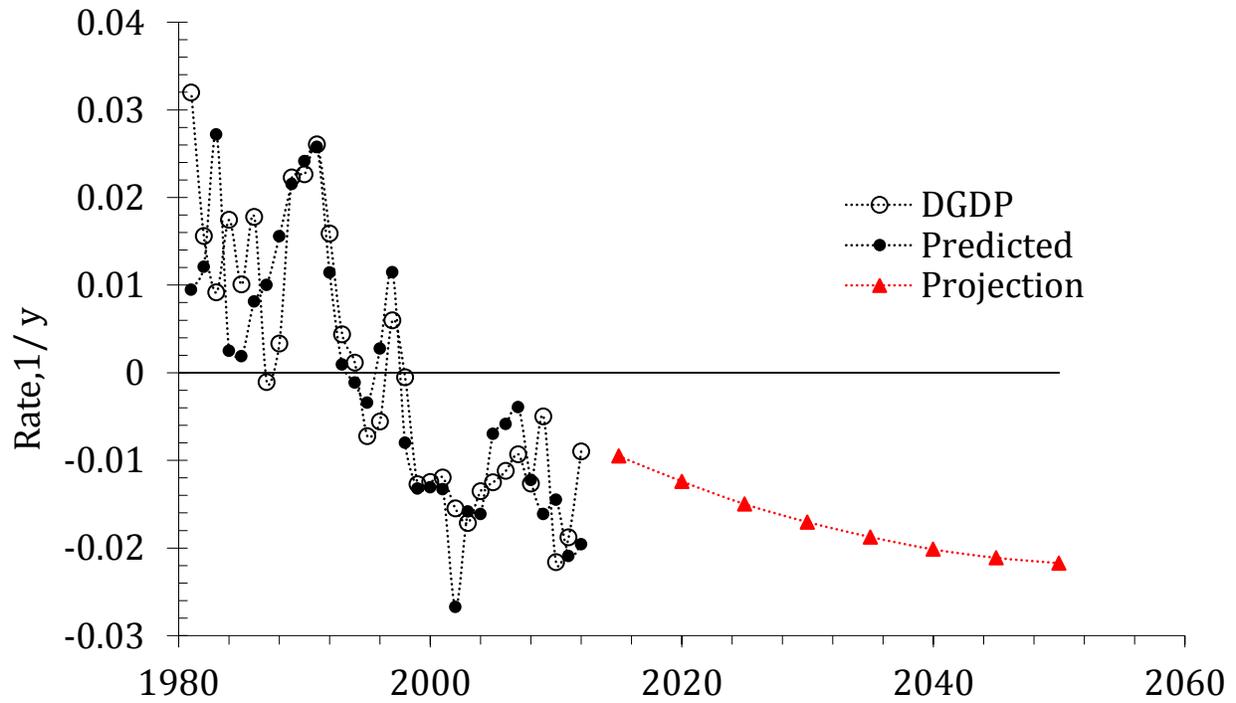

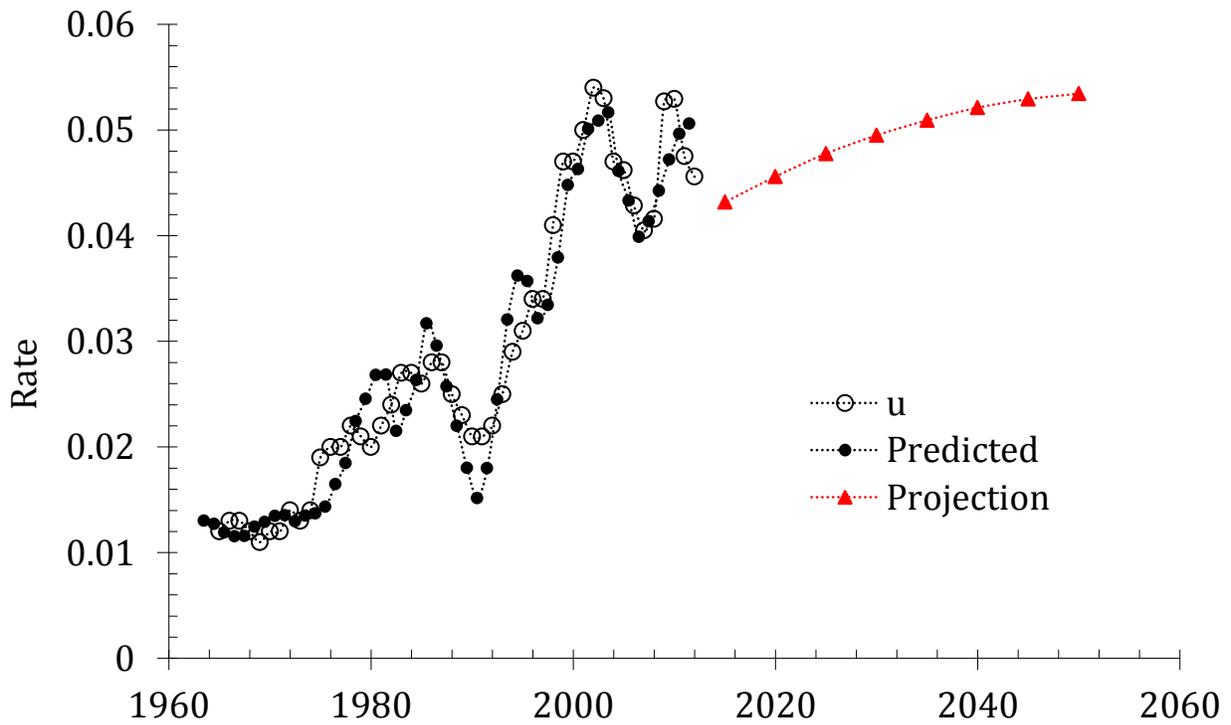

Figure 11. Prediction of the evolution of (upper panel) DGDP inflation rate, and (lower panel) the rate of unemployment through 2050.

**Conclusion**

There exists a Phillips curve for Japan in its original representation with a negative slope in the linear link between inflation and unemployment, with both variables evolving in sync. The existence of the Phillips curve does not facilitate the fight against deflation for the Bank of Japan. The deflationary period will last before the level of labor force will start to increase.

In Japan, the change rate of labor force level is the driving force behind unemployment and inflation. This finding confirms the existence of a generalized linear and lagged relationship between labor force, unemployment, and inflation in developed countries. The same relationship holds in the USA, France, Japan, Austria, Canada and Germany (Kitov and Kitov, 2010). The change in labor force in Japan does no lead inflation and unemployment. This observation differs from those in other developed countries, where time lags as large as 6 years are observed (Germany).

Labor force projections allow a reliable prediction of inflation and unemployment in Japan: DGDP inflation will be negative (between -0.5% and -1% per year) during the next 40 years. Unemployment will increase from ~4.0% in 2010 to ~5.5% in 2050

**References**

Ariga, K. and K. Matsu (2002). Mismeasurement of CPI, http://www2.e.u-tokyo.ac.jp/~seido/output/Ariga/ariga029.pdf.

Atkeson, A. and L.E. Ohanian (2001). Are Phillips Curves Useful for Forecasting Inflation? Federal Reserve Bank of Minneapolis Quarterly Review 25(1), 2-11.

Bullard, J., C. Garriga, C., and C. Waller (2012). Demographics, Redistribution, and Optimal Inflation, IMES Discussion Paper Series 12-E-13, Institute for Monetary and Economic Studies, Bank of Japan.

Bureau of Labor Statistic (2013). Foreign Labor Statistic. Table, retrieved at 20.07.2013 from http://data.bls.gov.

Caporale, G. M. and L. A. Gil-Alana (2006). Modeling structural breaks in the US, UK and Japanese unemployment rates, CESIFO Working Paper, 1734.

De Veirman, E. (2007). Which Nonlinearity in the Phillips Curve? The Absence of Accelerating Deflation in Japan, Reserve Bank of New Zealand, January 14, 2007.

Feyzioğlu, T. and L.Willard (2006). Does Inflation in China Affect the United States and Japan? IMF Working paper, WP/06/36.

Fisher, I. (1926). A Statistical Relation between Unemployment and Price Changes, *International Labor Review* 13: June, no. 6, 785-92. Reprinted as Irving Fisher. 1973. "I Discovered the Phillips Curve," *Journal of Political Economy* 81: March/April, no. 2, part I, 496-502.

Galí, J. and M. Gertler (1999). Inflation Dynamics: A Structural Econometric Analaysis, *Journal of Monetary Economics* 44, 195-222.

Gordon, K. (2009). The History of the Phillips Curve: Consensus and Bifurcation. *Economica* 78(309), 10-50.

Hayashi, N. (2005). Structural changes and unit roots in Japan's macroeconomic time series: is real business cycle theory supported? *Japan and the World Economy* 17, 239–259.

Ikeno, H. (2012). Convergence of Japanese Local CPIs with Structural Breaks, *河台経済論集, 21*(2).

Kamada, K. (2004). Real-Time Estimation of the Output Gap in Japan and its Usefulness for Inflation Forecasting and Policymaking, Deutsche Bank, Discussion Paper Series 1: Studies of the Economic Research Centre No 14/2004.

Kawata, H. and S. Naganuma (2010). Labor Force Participation Rate in Japan, Bank of Japan Review, 2010-E-7, December 2010.

Kitov, I. (2006). The Japanese economy, *MPRA Paper 2737*, University Library of Munich, Germany.

Kitov, I. (2009). The anti-Phillips curve, *MPRA Paper 13641*, University Library of Munich, Germany.

Kitov, I. and O. Kitov (2010). Dynamics Of Unemployment And Inflation In Western Europe: Solution By The 1- D Boundary Elements Method, *Journal of Applied Economic Sciences*, Spiru Haret University, Faculty of Financial Management and Accounting Craiova, 5(2(12)/Sum), 94-113.

Kitov, I. and O. Kitov (2012). Modeling Unemployment And Employment In Advanced Economies: Okun'S Law With A Structural Break, Theoretical and Practical Research in Economic Fields, Association for Sustainable Education, Research and Science, vol. 1(5), 26-41, June.

Leigh, D. (2004). Monetary Policy and the Dangers of Deflation: Lessons from Japan Department of Economics, Johns Hopkins University, August 10, 2004.

Lucas, R. E., Jr. (1972). Expectations and the Neutrality of Money, *Journal of Economic Theory* 4: April, no. 2, 103-24.
30

Lucas, R. E., Jr. (1973). Some International Evidence on Output-Inflation Tradeoffs, *American Economic Review* 63: June, no. 3, 326-34.

Nelson, E. (2006). The Great Inflation and Early Disinflation in Japan and Germany, Federal Reserve Bank of St. Louis, Working Paper 2006-052D, St. Louis.

Organization for Economic Cooperation and Development, (2005). Notes by Country: Japan. http://www.oecd.org/dataoecd/35/3/ 2771299.pdf.

Organization for Economic Cooperation and Development (2013). Corporate Data Environment, Labor Market Statistics, DATA, July 30, 2013.

Pascalau, R. (2007). Productivity shocks, unemployment persistence, and the adjustment of real wages in OECD countries, *MPRA Paper 7222*, University Library of Munich, Germany.

Phillips, A. W. (1958). The Relation between Unemployment and the Rate of Change of Money Wage Rates in the United Kingdom, 1861-1957, *Economica* 25, 283-299.

Rudd, J. and K. Whelan (2005a). New tests of the New Keynesian Phillips curve, *Journal of Monetary Economics,* vol. 52, pages 1167-1181.

Rudd, J. and K. Whelan (2005b). Modelling Inflation Dynamics: A Critical Review of Recent Research, Finance and Economics Discussion Series 2005-19, Board of Governors of the Federal Reserve System.

Sekine, T. (2001). Modeling and Forecasting Inflation in Japan, IMF Working Paper, WP/01/82.

Shirakawa, M. (2012). Demographic Changes and Macroeconomic Performance: Japanese Experiences, Opening Remarks, *Monetary and Economic Studies*, Bank of Japan, v. 30, November 2012.

Shiratsuka, S. (1999). Measurement errors in Japanese Consumer Price Index, Working Paper Series WP-99-2, Federal Reserve Bank of Chicago.
31

Sims, C. A. (2007). Monetary Policy Models, Brookings Papers on Economic Activity, Economic Studies Program, The Brookings Institution, vol. 38(2), 75-90.

Sims, C. A. (2008). Inflation Expectations, Uncertainty, the Phillips Curve, and Monetary Policy, paper presented at Federal Reserve Bank of Boston Conference, June 10-11. http://www.bostonfed.org/economic/conf/conf53/papers/Sims.pdf

Statistics Bureau, the Ministry of Internal Affairs and Communications (2013). Retrieved on July 30, 2013 from http://www.stat.go.jp/english/index.htm,.

Stock, J.H. and M.W. Watson (1999). Forecasting inflation, *Journal of Monetary Economics*, Elsevier, 44(2), 293-335.

Stock, J.H. and M.W. Watson (2002). Forecasting Using Principal Components From a Large Number of Predictors, *Journal of the American Statistical Association*, 97(460), 1167-1179.

Stock, J.H. and M.W. Watson (2006). Forecasting with many predictors, *Ch. 10 in Handbook of Economic Forecasting*, V.1, 515-554, Elsevier.

Stock, J.H. and M.W. Watson (2007). Why Has Inflation Become Harder to Forecast, *Journal of Money, Credit, and Banking* 39, 3-34.

Stock, J.H. and M.W. Watson (2008). Phillips Curve Inflation Forecasts. *Ch. 3 in* Understanding *Inflation and the Implications for Monetary Policy*, in J. Fuhrer, Y. Kodrzycki, J. Little, and G. Olivei, eds., Cambridge: MIT Press, 2009.

Stock, J.H. and M.W. Watson (2010). Modeling Inflation After the Crisis, *NBER Working Papers 16488*, National Bureau of Economic Research, Inc.
32